\documentclass[12pt]{iopart}
\usepackage{graphicx,epsfig}
\begin{document}
\title[Tunnelling induced collapse in time-dependent double-well potential]{Tunnelling induced collapse
 of an atomic Bose-Einstein condensate in a double-well potential}
\author{E. Sakellari, N.P. Proukakis and C.S. Adams}
\address{Department of Physics, University of Durham, South
Road, Durham DH1 3LE, United Kingdom}
  
\ead{n.p.proukakis@durham.ac.uk}

\begin{abstract}

The stability of a low temperature Bose-Einstein condensate with attractive
interactions in one and three dimensional double-well potentials is
discussed. In particular, the tunnelling dynamics of a
condensate under the influence of a time-dependent potential
gradient is investigated. The condensate is shown to collapse at
a critical potential gradient which corresponds to a critical
number of atoms in one of the two wells. The
sensitivity of this tunnelling induced collapse could provide a
useful tool in the study of condensates with attractive
interactions.

\end{abstract}

\maketitle

\section{Introduction}

The experimental realization of Bose-Einstein condensation (BEC) in
dilute atomic gases of $^{87}$Rb \cite{cornell}, $^{23}$Na \cite{ketterleNa},
$^{7}$Li \cite{hulet1}, $^{1}$H \cite{kleppner}, metastable $^{4}$He \cite{aspect},
 $^{85}$Rb \cite{cornish}, $^{41}$K \cite{inguscio}, $^{133}$Cs \cite{grimm} and
$^{174}$Yb \cite{japan} has stimulated theoretical studies of the
properties of ultracold Bose gases. The densities of such systems
are typically sufficiently low, that their interactions 
can be described by a single
parameter, the $s$-wave scattering length $a$. In addition 
one can control
not only the strength of these interactions but also
whether they are attractive ($a<0$ e.g. $^{7}$Li,
$^{85}$Rb, $^{133}$Cs at low magnetic fields) or repulsive
($a>0$) using magnetic-field induced
Feshbach resonances \cite{fesh}. Whereas a 
harmonically confined BEC with repulsive
interactions is stable for any number of atoms, a condensate with
attractive interactions is only stable if the atom number ${\cal N}$ is
smaller than a critical value ${\cal N}_{\rm cr}$
\cite{hulet1,ruprecht,dodd,houbiers,pitaevskii,kkagan,shuryak,dalf_str,review}. For
${\cal N}>{\cal N}_{\rm cr}$ the interaction energy exceeds the zero-point
kinetic energy and a collapse occurs, as extensively studied in
\cite{sstoof,perez_garcia,shi_zheng,ffetter,leggett,parola,kagan,huang,adhikaricyl,gammal1}.
 In experiments on $^{7}$Li, the magnetic field is
held fixed and the number of condensate atoms grows up to ${\cal N}_{cr}$
where a partial collapse occurs \cite{hulet5,hulet6}. In this system, it should
be noted that the attractive interactions can lead to the formation of bright matter-wave soliton trains 
 in elongated optical traps
\cite{huletsoliton}.
In contrast to the $^{7}$Li experiments, in the case of
$^{85}$Rb, a Feshbach
resonance \cite{cornish} has been used to switch the
scattering length from positive to negative values, producing a
condensate with ${\cal N}\gg {\cal N}_{cr}$ which subsequently collapses
\cite{donley}, as modelled in 
\cite{duine,adhikari2,santos,saito,savage,hollandnjp,adhikarijpb}

% There have been many theoretical estimates for
% ${\cal N}_{\rm cr}$ both numerically using mean-field calculations
% \cite{ruprecht,adhikaricyl,gammal1} and using variational
% methods \cite{review,fetter,leggett}. Several authors have modelled
% $^{85}$Rb collapse experiments, by numerical solution of the
% Gross-Pitaevskii (GP) equation
% \cite{adhikari2,santos,saito,savage,hollandnjp,adhikarijpb,bao} or using a variational method \cite{duine}.
% In experiments on $^{7}$Li, the magnetic field is
% held fixed and the number of condensate atoms grows up to ${\cal N}_{cr}$
% where a partial collapse occurs \cite{hulet5,hulet6}. More
% recently, the formation of bright matter-wave soliton trains in
% $^{7}$Li atoms in elongated optical traps has been reported
% \cite{huletsoliton}.

Although much work has been done to study the collapse properties
of dilute BECs in a single-harmonic trap, an interesting subject
is the behaviour of a condensate with attractive
interactions in a double-well potential. Magnetic
\cite{mit_dw,magnetic_dw} and optical
\cite{mit_interferometer_2,distillation} double-well potentials
have been created in recent experiments and a proposal for a
magnetic double-well has been reported \cite{foot}. Although
condensates with $a>0$ in a double-well system have received considerable 
theoretical attention \cite{Two_State_0,Two_State_2,shen,Two_State_3,
Two_State_5,MF_3,fant,sak1,sak2}), only Adhikari \cite{adhikari} 
and Coullet {\it et al.} \cite{coullet} have studied the $a<0$ case.

In this paper, we investigate the stability of a low temperature atomic
condensate with attractive interactions in a double-well
potential, where the Josephson dynamics are induced by the
addition of a time-dependent potential gradient. 
Our analysis is based on the Gross-Pitaevskii equation. The aim
is to discuss the stability of the condensate in the double-well potential
and show how tunnelling between the wells can lead to collapse. We do not, 
however, intend to describe the {\it collapse dynamics}, for which, one should at least
additionally incorporate a suitable 3-body loss term \cite{saito,bao}.

Section 2 introduces the main formalism and briefly reviews the
eigenenergies of the double-well potential as a function of the
potential gradient in both one and three dimensions. Section 3
discusses the stability of a BEC in a three dimensional symmetric
double-well as a function of the barrier height. Finally, Section 4 
considers the time-dependent evolution of a 3D system as the
gradient is increased at a constant rate. We find that a
collapse occurs as the gradient is increased above a
critical value, which, within our formalism, can be predicted 
by the eigenstate curves.

\section{Eigenenergy levels of a BEC in a double-well potential}

At low temperatures, the condensate wavefunction
$\psi\left(\mbox{\boldmath$r$},t\right)$ (normalized to unity)
obeys the dimensionless Gross-Pitaevskii (GP) equation,
\begin{eqnarray}
i \frac{\partial}{\partial t}\psi \left( \mbox{\boldmath$r$},t\right) =  \left[-\frac{1}{2} \nabla^{2} +V\left(
\mbox{\boldmath$r$}\right)+g_{\rm
3D}|\psi(\mbox{\boldmath$r$},t)|^{2}\right] \psi
(\mbox{\boldmath$r$},t)~. \label{eq:GP3dhou}
\end{eqnarray}
In the above equation, $V\left(\mbox{\boldmath$r$}\right)$ is the
confining potential and $g_{\rm 3D}=g/(a_{\perp}^3 \hbar
\omega_\perp)$, is the dimensionless effective interaction term,
where $g={\cal N}(4\pi\hbar^2a/m)$ is the usual three-dimensional
scattering amplitude, defined in terms of the {\it s}-wave
scattering length $a$, the total number ${\cal N}$ of atoms
of mass $m$, and  the
harmonic oscillator length in the transverse direction(s) $a_{\perp}=\sqrt{\hbar/m\omega_{\perp}}$,
 with $\omega_{\perp}$ the corresponding trapping frequency. The
confining potential,
\begin{equation}
V\left({\bf r}\right)=\frac{1}{2}\left( \rho^{2} +\lambda^2 z^2\right)+%
h\exp\left(-z^2\right)+\delta z~. \label{eq:conf3d}
\end{equation}
describes an axially symmetric potential,
with $\rho^{2}=x^{2}+y^{2}$ and asymmetry parameter $\lambda=\omega_{z}/\omega_{\perp}$ plus
a Gaussian potential of height $h$, located
at $z=0$. In addition, a linear potential $\delta z$ of gradient
$\delta$ is pivoted at the centre of the trap. For $\delta>0$, the
right well has higher potential energy and the trap centre is
additionally shifted into the $z>0$ region; however, this shift is
negligible
 for the parameters studied throughout this work. Throughout this paper we work in
dimensionless (harmonic oscillator) units, by applying the
following scalings: space coordinates
 transform according to
$\mbox{\boldmath$r$}_{i}^{\prime}={a_{\perp}}^{-1}\mbox{\boldmath$r$}_{_{i}}$, time $t^{\prime}=\omega_{\perp} t$,
condensate wavefunction
$\psi^{\prime}\left(\mbox{\boldmath$r$}^{\prime},t^{\prime}\right)=\sqrt{a_{\perp}^3}\psi\left( \mbox{\boldmath$r$},t\right)$
  and energy $E^{\prime}=\left(\hbar\omega_{\perp}\right)^{-1}E$ (primes henceforth neglected for convenience).

We can also consider the 1D form of Eq.~(\ref{eq:GP3dhou}), which
accurately takes into account the transverse dynamics of `cigar'
condensates ($\lambda<1$),
\begin{eqnarray}
i \frac{\partial}{\partial t}\psi \left( z,t\right) =  \left[ -\frac{1}{2} \frac{\partial^2}{\partial z^2}%
 +V_{\rm 1D}\left(z\right)+g_{\rm
1D}|\psi(z,t)|^{2}\right] \psi (z,t)~, \label{eq:GP1dhou}
\end{eqnarray}
where $\psi (z)$ is normalized to unity, $g_{\rm 1D}=g_{\rm
3D}/(2\pi a_{\perp}^2)$ is the 1D self-interaction parameter that
matches the 1D and 3D axial density profiles and $V_{\rm 1D}(z)$ is the confining potential in
the axial direction.
\begin{equation}
V_{\rm 1D}\left(z\right)=\frac{1}{2}z^2+h\exp\left(-z^2\right)+\delta z~. \label{eq:conf1d}
\end{equation}

\begin{figure}[hbt]
\centering
\includegraphics[width=14.0cm]{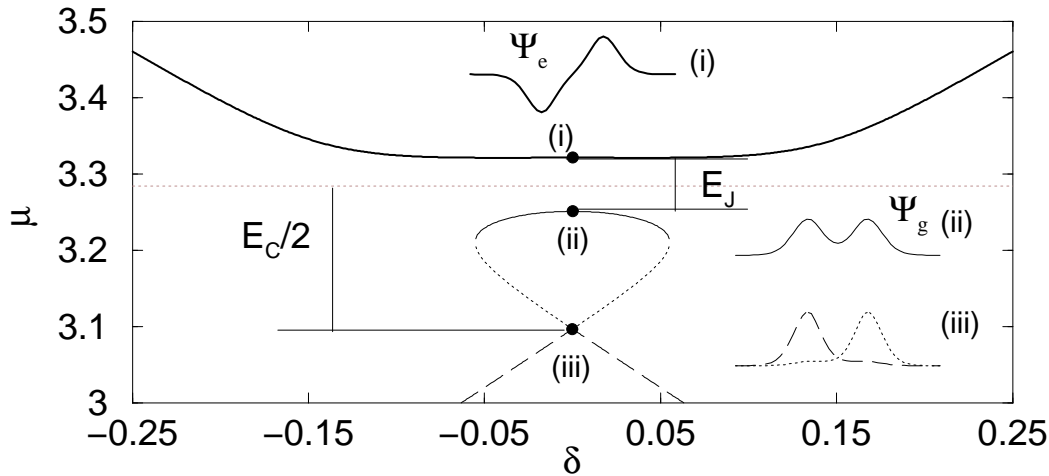}
\caption{ Eigenenergies $\mu~(\hbar\omega_{\perp})$ for the double-well for an attractive Bose gas as a function of the
potential gradient $\delta~(\hbar \omega_{\perp}/a_{\perp})$ indicating the self-interaction energy,
 $E_{\rm C}$, and the Josephson
 coupling energy, $E_{\rm J}$ in each case. The horizontal
dotted grey line corresponds to the zero energy of the two-state model.
The value of the nonlinearity used here is $g_{\rm 3D}=-\pi$, corresponding to $E_{\rm C}=-0.379
\hbar\omega_\perp $ and $E_{\rm J}=0.071 \hbar\omega_\perp$. The
eigenstates at the centre of the trap are also shown in each case.
We assume a spherical trap geometry ($\lambda=1$),
and a Gaussian barrier of height $h=4 \hbar \omega_{\perp}$ located at the centre of the trap.}
\end{figure}

The eigenstates of the double-well condensate in 1D and 3D are
 calculated by substituting $\psi\left(\mbox{\boldmath$r$},t\right)=
e^{-i\mu t}\Psi \left(\mbox{\boldmath$r$}\right)$, where $\mu$ is
the chemical potential (of the one and three dimensional system
respectively), and solving the resulting time-independent equation
as discussed in \cite{sak1}. For small nonlinearities there are
only two levels and, for $\delta=0$, the eigenstates are a symmetric
ground state $\Psi_g$ and an antisymmetric first excited state,
$\Psi_e$, with equal population in both wells. Sufficiently large
interactions lead to the appearance of a loop structure (see e.g.
\cite{Two_State_5,Loop}). This loop structure appears in the first excited state for $a>0$, and in
the ground state for $a<0$.
In this paper we consider the latter case (Fig. 1), for which
the corresponding wavefunctions at $\delta=0$ are: (i) a
symmetric ground state $\Psi_{g}$ with equal population in both
wells, (ii)  an anti-symmetric
state with equal population in both wells and a phase difference
of $\pi$ across the trap centre, which we shall henceforth refer
to as $\Psi_{e}$,
and (iii) two lower energy, non-symmetric ground state solutions
with most of the particles in either the left (dashed) or the
right (dotted) well (the so-called `self-trapped' states
\cite{Two_State_2,shen,sak1,sak2}).

Under certain conditions, the eigenenergies $\mu$ can also be reproduced by the two-state
model (see e.g.
\cite{Two_State_0,Two_State_2,shen,Two_State_3,Two_State_5}) described by the Hamiltonian,
\begin{equation}
H=\frac{1}{2}\left(\begin{array}{cc}
-\Delta+E_{\rm C}N & -E_{\rm J} \\
-E_{\rm J}        & \Delta-E_{\rm C}N
\end{array}\right)~,
\label{eq:hamiltonia}
\end{equation}
where $N= (N_\ell-N_r) /{\cal N}$ is the fractional population
difference, $\Delta$ is the potential energy difference
 between the left ($\ell$) and right ($r$) wells, $E_{\rm C}=g\langle\Psi_{\ell,r} |
|\Psi_{\ell,r}|^{2}|\Psi_{\ell,r}\rangle$ is the self-interaction
energy, $g$ is the nonlinearity and $E_{\rm
J}=-2\langle\Psi_\ell\vert\left(-\textstyle{1\over2} \nabla
^2+V_{\delta=0}\right)\vert\Psi_r\rangle$ is the Josephson
coupling energy. The energy splittings $E_{\rm C}$ and $E_{\rm J}$
are indicated in Fig.~1. For the case
of a potential gradient $\delta$, we can write
$\Delta=\alpha\delta$, where
 $\alpha$ is a numerical factor determined numerically from the GP solution.
The picture for the eigenenergy levels as a function of $\delta$
shown in Fig.~1, is only valid in 1D and 3D for a range of
nonlinearities. 
%%%To
%%%illustrate the behaviour for the complete range of nonlinearities
%%%we plot the value of splittings $|E_{\rm C}|$ and $E_{\rm J}$ as a function of
%%%$g_{\rm 1D}$ and $g_{\rm 3D}$ in Figs.~2 and 3 respectively.

\begin{figure}[hbt]
\centering
\includegraphics[width=14.0cm]{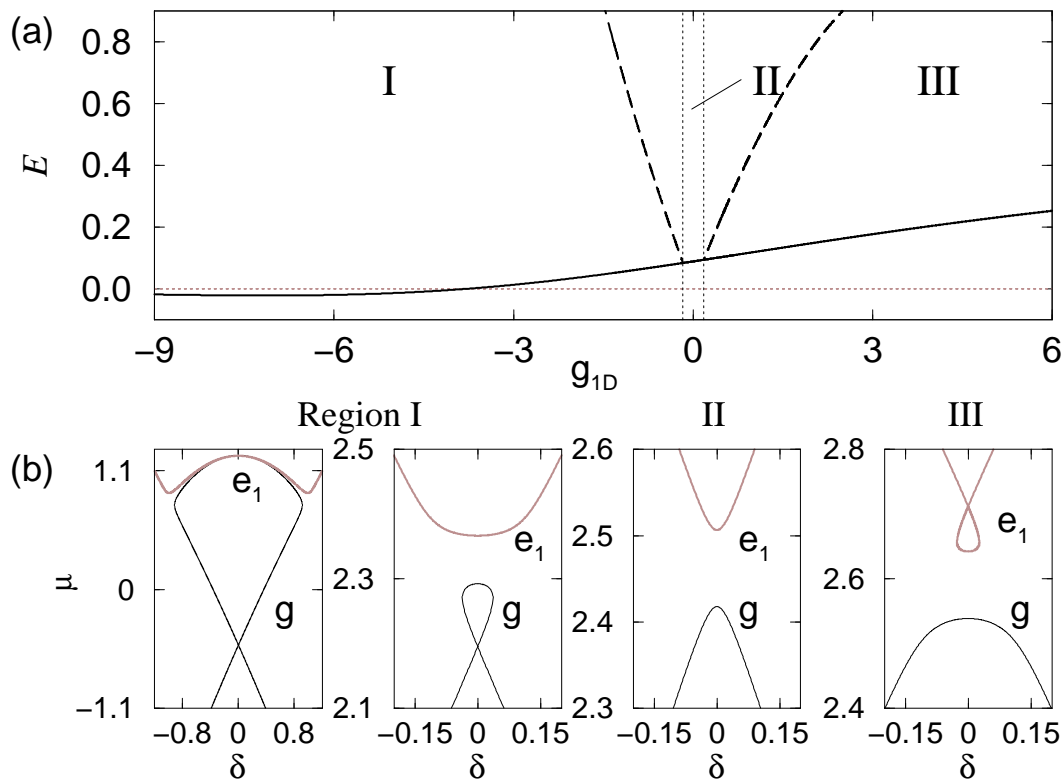}
\caption{(a) The self-interaction energy $|E_C|$ (dashed lines) and the
 Josephson energy $E_J$ (solid) at $\delta=0$ as a function of
 the nonlinearity $g_{\rm 1D}$ for the 1D confining trap with $h=4\hbar \omega_{\perp}$.
 (b) Typical eigenenergies $\mu$ as a function
of the potential gradient $\delta$ for various nonlinearities within regions I-III.
Shown are (from left to right) the cases
 $g_{\rm 1D}=-4$, $-0.5$, $0$ and $0.5$. For $g_{\rm 1D}<-3.79$, $E_{\rm J}<0$. The horizontal
dotted grey line corresponds to $E=0$.}
\end{figure}

Focusing first on the 1D case, we plot in
 Fig.~2(a) the two-state model parameters $|E_{\rm C}|$
 and $E_{\rm J}$ at $\delta=0$ (symmetric double-well) as a function of $g_{\rm 1D}$.
 For $g_{\rm 1D}<0$,
increasing the magnitude of the nonlinearity leads to the
appearance of a loop structure in the ground state at a critical
point $|E_{\rm C}|=E_{\rm J}$. As a 1D condensate is stable
against collapse \cite{hulet1,ruprecht,shuryak,kagan,review,huang,adhikaricyl}, $|E_{\rm C}|$ remains
finite and a loop structure is always observed. Note that, for a
nonlinearity less than a critical value, the splitting $E_{\rm J}$ becomes negative,
 signifying
an inversion of the lowest two energy eigenvalues. This can be
explained using $\mu=E-1/2|g|\int |\Psi|^4dr=E-|E_{\rm int}|$.
Although $E$ is larger for the first excited state than for the
ground state, the first excited state has more negative
interaction energy (as its peak density is higher), thereby reducing
$\mu$ below the ground state value.
Fig.~2(b) shows typical eigenenergy levels for the ground and
the first excited states as a function of the potential gradient
for different values of the nonlinearity. The two-state model
correctly reproduces the eigenenergy curves for small values of
$\vert g_{\rm 1D}\vert$ but cannot reproduce the
inversion of the eigenenergy levels for $g_{\rm 1D}\ll 1$. The
case $g_{\rm 1D}>0$ has been discussed in our earlier work \cite{sak1}.

In Fig.~3 we consider the 3D case. In Fig.~3(a) we plot the
splittings $|E_{\rm C}|$ and $E_{\rm J}$ at $\delta=0$ as a
function of $g_{\rm 3D}$. In contrast to 1D, for $g_{\rm 3D}<0$
the condensate collapses when the atom number, or magnitude of the nonlinearity,
exceeds a critical value. The collapse appears first
in the self-trapping states. In Fig. 3(a) this corresponds to the
point where the curve for $|E_{\rm C}|$ (dashed line in Fig.~3(a)) terminates at the boundary between region I and II 
(indicated by the vertical dotted line in Fig.~3(a)). At larger negative nonlinearities (region I in Fig. 3(a)) the lowest eigenstates
 invert (as in the 1D case) and at $\delta=0$ the symmetric states also become unstable at $g_{\rm 3D}=-11.6$. In Fig.~3(b)
we plot typical eigenenergy levels as a function of $\delta$. The
curves are similar to the 1D case except in the limit of large
negative nonlinearities (region I), where a completely different
structure is found. Note that region I cannot be described by the two-state model. In this region there is no longer a loop
structure as the self trapped states are unstable. This parameter
region is of interest for investigating tunnelling induced
collapse, where one begins with a stable symmetric state and adds a
potential gradient to induce a collapse in one well. However, before discussing the dynamical
behaviour we consider the stability in the symmetric double-well
as a function of the barrier height.

\begin{figure}[t]
\centering
\includegraphics[width=14.0cm]{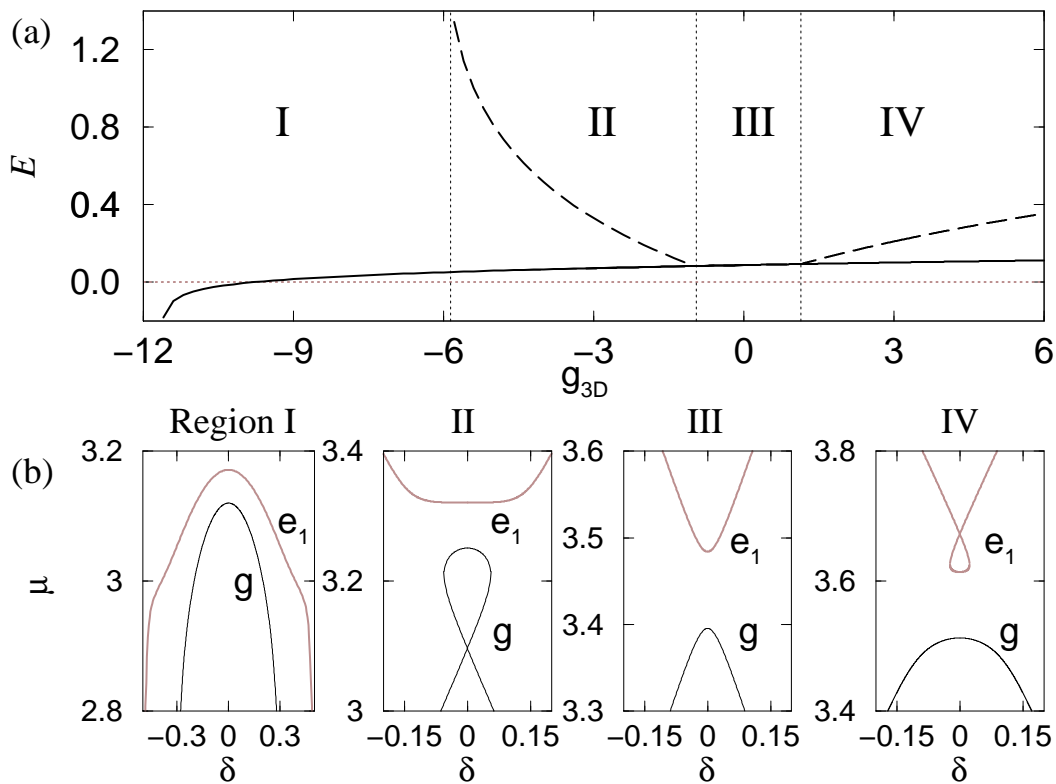}
\caption{(a) As in Fig. 2, but for the 3D case, with $\lambda=1$ and $h=4\hbar\omega_{\perp}$.
(b) Typical eigenenergies shown here are for each of the four regions I-IV in (a),
namely (from left to right)  $g_{\rm 3D}=-6,-\pi,0$ and $\pi$. Compared to the 1D case (Fig. 2), 
in 3D there is an additional region (I) corresponding to the case 
where the self-trapped states become unstable. The horizontal
dotted grey line corresponds to $E=0$.}
%The self-interaction energy $|E_C|$ (dashed) and the
% Josephson energy $E_J$ (solid) lines at $\delta=0$ as a function of the nonlinearity $g_{\rm 3D}$
% for the 3D confining trap with $\lambda=1$ and $h=4\hbar
% \omega_{\perp}$. 
% Typical eigenenergies $\mu$ as a function of the
% potential gradient $\delta$ for various nonlinearities (from left
% to right) $g_{\rm 3D}=-6,-\pi,0$ and $\pi$. 

\end{figure}

\section{Stability of a 3D BEC in a double-well potential}

In this Section we study the stability of a BEC with attractive
interactions in a 3D symmetric double-well trap
as a function of the barrier height $h$.
We solve Eq.~(\ref{eq:GP3dhou}) numerically
using the Newton method \cite{sak1,numrec}.
Above the critical atom number ${\cal N}_{\rm cr}$
we no longer find stationary solutions.
A dimensionless constant $k_{\rm cr}$ relating
the scattering length $a$ with ${\cal N}_{\rm cr}$ and
the properties of the confining trap, is defined by, \cite{ruprecht}
\begin{equation}
\frac{{\cal N}_{cr} |a| }{a_0}=k_{cr}~, \label{eq:criticalparam}
\end{equation}where $a_0=\sqrt{\hbar/m\omega_0}$, $m$ is the mass of the atoms confined in the trap and
$\omega_0=\lambda ^{1/3} \omega_{\perp}$ is the geometrically
 averaged trap frequency.

\begin{figure}[t]
\centering
\includegraphics[width=12.0cm]{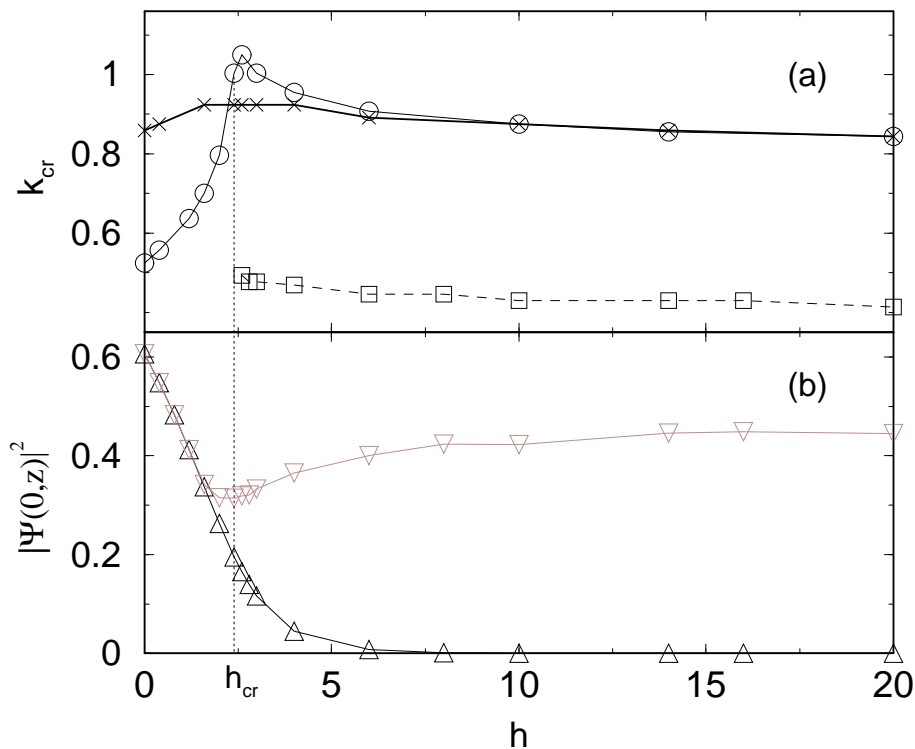}
\caption{(a) Critical parameter $k_{cr}$ as a function of $h$ for
a spherical geometry ($\lambda=1$). Stability curves of the
symmetric $\Psi_g$ (curve (ii) in Fig. 1), first excited $\Psi_e$ (curve (i)) and asymmetric ground
states (iii) are shown as circles, crosses and squares respectively. (b) Condensate density $|\Psi(0,z)|^2$ along
the $z$ axis as a function of
 $h$ for the state $\Psi_g$. Shown are the central $|\Psi(0,0)|^2$ (black) and peak $|\Psi(0,z_m)|^2$ (grey)
 densities, where $z_m$ is the longitudinal
  position of maximum density in the double-well configuration (i.e. centre of each 
individual well). In both figures, the plotted lines connect adjacent data points.
The vertical dotted line highlights the critical value $h_{cr}$, above which the
stability curve consists of two branches.}
\end{figure}

In Fig.~4(a) we plot the critical
 constant $k_{cr}$ as a function of $h$ for the case of a
symmetric double-well trap with $\lambda=1$ for the ground and first excited states.
 We find that at a critical
 value of the barrier height, $h_{cr}$, there are two branches 
to the stability curve for the ground state. The upper and lower branches correspond to
the symmetric (Fig. 1, curve (ii)) and asymmetric (Fig. 1, curve (iii))
eigensolutions respectively, (plotted by circles and squares in Fig. 4). Note also that $k_{\rm cr}$
reaches its maximum value at a height just above that corresponding
to the appearance of the second branch. This maximum can be
explained by a minimum in the peak density of the double-well configuration,
which is plotted in
Fig. 4(b). As the barrier is raised the condensate splits in
two, thus reducing its maximum density. However, as the trap splits to form
two separate condensates, the condensates in each well become
compressed and the peak density increases again. 
Our results remain qualitatively unchanged for
different values of $\lambda$.
This picture is 
similar to that of Adhikari \cite{adhikari}, except for the behaviour at large $h$,
where Adhikari finds that $k_{\rm cr}$ increases again. 
As anticipated, we find that, for $h>h_{cr}$, when the system is essentially
composed of two separate condensates, the value
of $k_{\rm cr}$ in each well tends towards the value in a single harmonic trap containing the same number of atoms
as each half of the double-well.

Finally, the critical constant $k_{\rm cr}$ as a function of $h$, for 
the first excited state, $\Psi_{\rm e}$ (crosses in
Fig. 4(a)) first increases, reaches a peak value, and then decreases
approaching the value for the symmetric ground state for large $h$. 
This is expected, as in the limit of large $h$, the density distributions
of $\Psi_{\rm g}$ and $\Psi_{\rm e}$ become very
similar.

\section{Tunnelling Dynamics under a time-dependent magnetic field gradient}

The main theme of this Section is to investigate the possibility
of observing a tunnelling induced collapse.
At $t<0$ we prepare a stable condensate in a symmetric double-well, for a value of the
nonlinear constant $g_{\rm 3D}$ in region I of Fig.~3(a). 
In order to induce a collapse, a potential gradient is applied
at $t=0$, i.e., $\delta=Rt$ for $t>0$, such that the right well
has higher potential energy than the left. Subsequently, we
study the dynamics leading to the onset of collapse, by solving
Eq.~(\ref{eq:GP3dhou}) numerically. Note that this qualitative picture should remain correct,
even if 3-body loss terms are included in the treatment, although the latter may
affect the precise value for the onset of the collapse.

\begin{figure}[t]
\centering
\includegraphics[width=13.0cm]{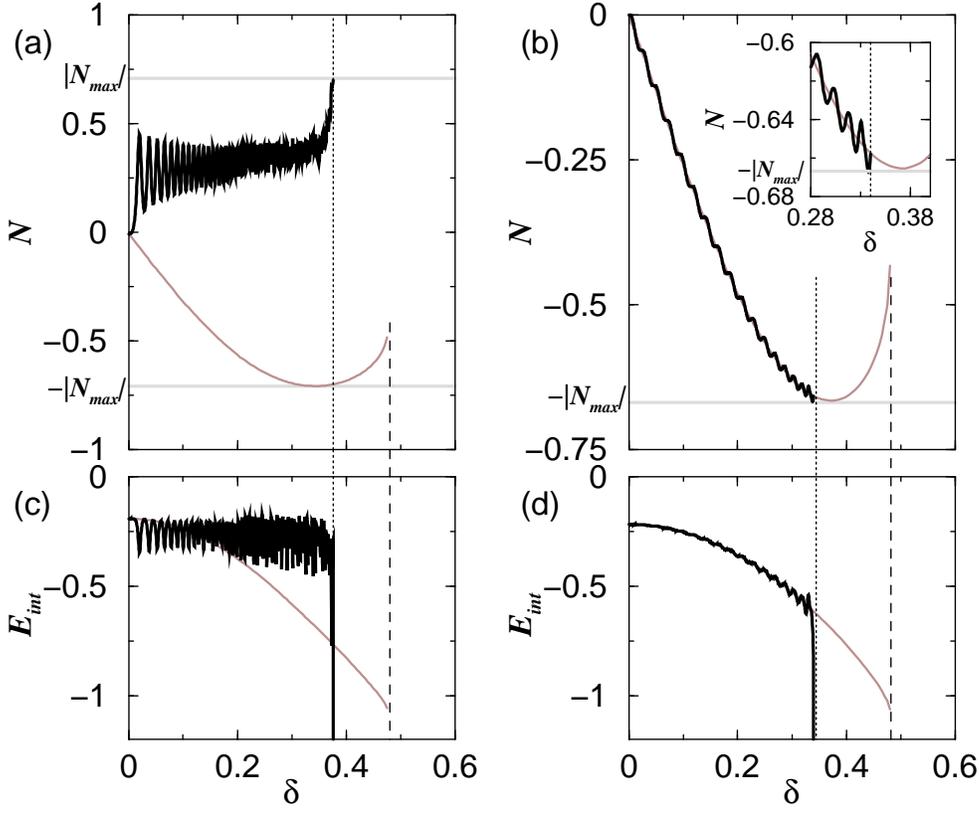}
\caption{ 
(a)-(b) Evolution of fractional population difference $N$ as a function 
of $\delta$ for a system initially prepared in state (a) $\Psi_g$ and (b) $\Psi_e$ (black lines). 
The population
 difference for the eigenstates are also shown as solid grey lines. The vertical dashed lines mark the critical gradient 
at which the eigenstates become unstable ($\delta_0=0.475$ and $\delta_0^{\prime}=0.480$ for $\Psi_g$ and $\Psi_e$ respectively).
 The vertical dotted lines describe where the system collapses in the time-dependent 
simulation. The collapse occurs when $|N|$ reaches the maximum value $|N_{\rm max}|$ (indicated by horizontal grey lines in 
(a) and (b)) of the number asymmetry predicted by the eigenstates. (c)-(d) Evolution of the interaction  energy $E_{\rm int}$
 (thick black line) when the potential gradient $\delta=Rt$ is 
increased at a constant rate $R$ for (c) $\Psi_g$ and (d) $\Psi_e$, with
corresponding eigenenergies shown by grey lines. Other parameters
 used here: $g_{\rm 3D}=-7$, $h=4\hbar\omega_{\perp}$, $\lambda=1$ and $R= 10^{-3} \hbar\omega_{\perp}^{2}/a_{\perp}$.} 
\end{figure}

For attractive interactions, the population difference induced
by the addition of the gradient does
not follow that of the eigenstate, as shown in Fig. 5(a).
The effect of the nonlinearity is that the ground
state is immediately projected onto a superposition
of states. As the potential gradient is 
increased, the population in one well reaches a critical
value and a collapse occurs. The critical 
gradient corresponding to the collapse, shown
by the dotted vertical line in Fig. 5, is identified as
the point where the interaction energy 
$(|E_{\rm int}|=1/2|g_{\rm 3D}|\int |\Psi|^4dr)$ 
diverges, see Fig. 5(b).  Note that the critical value
of the number asymmetry $N$ for which the time-dependent collapse occurs is close to the maximum 
value of $|N_{\rm max}|$ found for the eigensolution (grey horizontal 
line in Fig. 5(a)). 

This prediction becomes clearer if we consider a 
condensate prepared in the first excited state $\Psi_e$,
where the time evolution closely follows the eigensolution,
see Fig. 5(b). In this case the collapse occurs at exactly the
point where the number asymmetry exceeds $|N_{\rm max}|$, see inset of Fig. 5(b).
The critical gradient at which the
collapse is observed is found to be essentially independent
of the rate $R$ at which the gradient is increased. 

One can also compare the critical number needed in
one well before collapse occurs with the prediction
for the symmetric potential shown in Fig. 4. By
defining $K_{\rm cr}={\cal N}'_{\rm cr}\vert a\vert/a_0$,
where ${\cal N}'$ is the number of atoms in the well which collapses 
we find that $K_{\rm cr}=0.471$ and 0.467 for ground and excited
states in Fig. 5, which is close to the value of $k_{\rm cr}=0.470$ predicted by the lower branch of Fig. 4.

\begin{figure}[hbt]
\centering
\includegraphics[width=14.0cm]{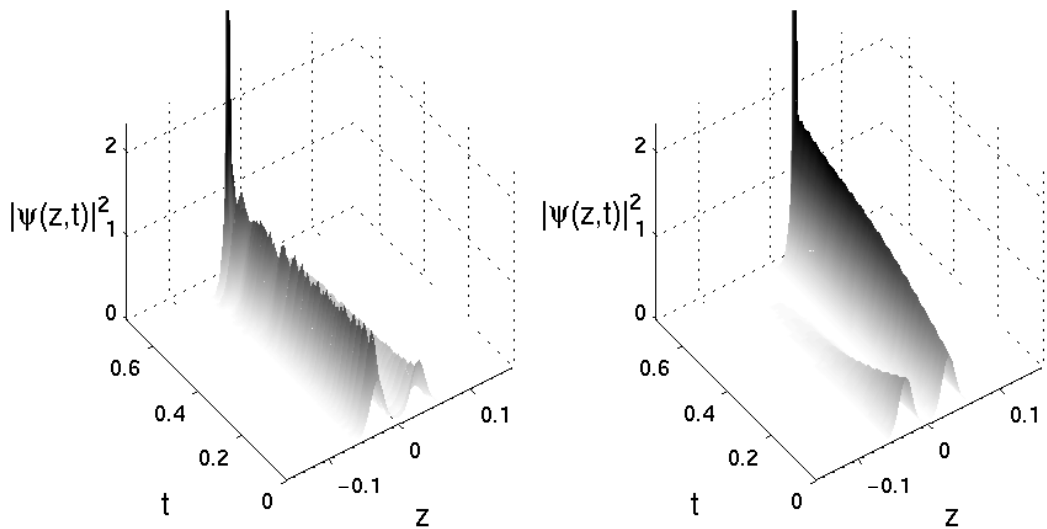}
\caption{Surface plot of the evolution of the density distribution $(|\Psi(z,t)|^2 \times 10^{10} {\rm cm^{-3}})$
along the $z$-axis $({\rm mm})$ as a function of time (s) for a BEC initially
prepared in the $\Psi_g$ (left) and $\Psi_e$ (right) states with
$g_{\rm 3D}=-6$. Due to the potential gradient, tunnelling is
induced to the left ($z<0$) well for the $\Psi_g$ state and to the right
well for $\Psi_e$. The condensate instability occurs at
$t=0.65$ s for $\Psi_g$ and $0.67$ s for $\Psi_e$. Other
parameters as in Fig.~5.}
\end{figure}

Finally, we discuss typical experimental parameters required for the demonstration of the tunnelling induced collapse. 
In the harmonic oscillator units discussed in Section~2, the number of atoms is given by, 
\begin{equation}
{\cal N}={\frac{g_{\rm 3D}}{4\pi}}{\frac{a_{\perp}}{a}}=
{\frac{g_{\rm 3D}}{4\pi a}}\sqrt{{\frac{\hbar}{m\omega_{\perp}}}}~.
\label{eq:numberpart}
\end{equation}
For $^7$Li atoms and taking $g_{\rm 3D}=-6$ and 
$\omega_{\perp}= 2 \pi \times 100$ Hz, we find
${\cal N}=1200$ which is below the critical value for collapse.
For an applied field gradient $R=(10^{-3})(\hbar \omega_{\perp}/a_{\perp})$ the collapse occurs at $t_{\rm exp} \sim 0.6$~s.
 The
collapse is illustrated by the density plots shown in Fig. 6.
We have confirmed that the collapse can be avoided
if the potential gradient is ramped up to a value smaller than the
critical gradient and then held constant.
This would not hold if the system were very close to the critical region,
in which case number fluctuations \cite{number1,number2} could enhance tunnelling and hence
induce the collapse at a slightly smaller gradient than that predicted by our simple model.

\section{Conclusions}

We have studied the stability of a low temperature atomic BEC with attractive interactions
in a 1D and 3D double-well potential. In particular we highlight a regime
where the condensate is stable if the population in both wells
are approximately equal, but becomes
unstable if there is sufficient tunnelling
from one well to the other. We
study the dynamics of the system when driven by
a time-dependent potential gradient and show
that a collapse occurs at a critical gradient predicted by the
time-independent solutions.
Although this picture is expected to be qualitatively correct in low temperature
atomic condensates, further work is required to determine the precise details of
the onset of collapse and the subsequent collapse dynamics, as well as the effects of
fluctuations \cite{number1,number2} and finite temperature \cite{thermal1,thermal2}.

\ack We acknowledge funding from the UK EPSRC.

%\newpage


\vspace{2.0cm}
\section*{References}
\begin{thebibliography}{99}

\bibitem{cornell} Anderson M H, Ensher J R,
Mathews M R, Wieman C E, and Cornell E A, Science {\bf 269},
198 (1995)

\bibitem{ketterleNa}
Davis K B, Mewes M -O, Andrews M R, van Druten N J, Durfee D S, Kurn D M, and Ketterle W, Phys. Rev. Lett. {\bf 75},
3969 (1995)

\bibitem{hulet1}
Bradley C C, Sackett C A, Tollett J J, and Hulet R G, Phys.
Rev. Lett. {\bf 75}, 1687 (1995); Bradley C C, Sackett C A,
and Hulet R G, Phys. Rev. Lett. {\bf 78}, 985 (1997)

\bibitem{kleppner}
Fried D G, Killian T C, Willmann L, Landhuis D,  Moss S C,
kleppner D, and Greytak T J, Phys. Rev. Lett. {\bf 81}, 3811
(1998)

\bibitem{aspect}
Robert A, Sirjean O, Browaeys A, Poupard J, Nowak S, Boiron D, Westbrook C I, and Aspect A,
 Science {\bf 292}, 461 (2001)

\bibitem{cornish}
Cornish S L, Claussen N R, Roberts J L, Cornell E A and Wieman C E, Phys. Rev. Lett. {\bf 85}, 1795 (2000)

\bibitem{inguscio}
Modugno G, Ferrari G, Roati G, Brecha R J, Simoni A, and Inguscio M, Science {\bf 294}, 1320 (2001)

\bibitem{grimm}
Weber T, Herbig J, Mark M, Nagerl H -C, and Grimm R, Science {\bf 299}, 232 (2003)

\bibitem{japan}
Takasu Y, Maki K, komori K, Takano T, Honda K, Kumakura M, Yabuzaki T, and Takahashi Y,
Phys. Rev. Lett. {\bf 91}, 040404 (2003)

\bibitem{fesh}
Tiesinga E, Verhaar B J, and Stoof H T C, Phys. Rev. A {\bf
47}, 4114 (1993); Courteille P, Freeland R S, Heinzen D J, van Abeelen F A, and Verhaar B J, Phys. Rev. Lett. {\bf 81}, 69
(1998); Inouye S, Andrews M R, Stenger J, Miesner H -J, Stamper-Kurn D M and Ketterle W, Nature {\bf 392}, 151 (1998)

%---------------------------------------------------------------------

\bibitem{ruprecht}
Ruprecht P A, Holland M J, Burnett K, and Edwards M, Phys.
Rev. A {\bf 51}, 4704 (1995)

\bibitem{dodd} Dodd R J, Edwards M, Williams C J, Clark C W, Holland M J, Ruprecht P A and Burnett K,
Phys. Rev. A {\bf 54}, 661 (1996)

\bibitem{houbiers} Houbiers M and Stoof H T C, Phys. Rev. A {\bf 54}, 5055 (1996)

\bibitem{pitaevskii} Pitaevskii L P, Phys. Lett. A {\bf 221}, 14 (1996)

\bibitem{kkagan} Kagan Yu, Shlyapnikov G V and Walraven J T M, Phys. Rev. Lett. {\bf 76}, 2670 (1996)

\bibitem{shuryak}
Shuryak E V, Phys. Rev. A {\bf 54}, 3151 (1996)

\bibitem{dalf_str} Dalfovo F and Stringari S, Phys. Rev. A {\bf 53}, 2477 (1996)

\bibitem{review}
Dalfovo F, Giorgini S, Pitaevskii L P, and Stringari S, Rev.
Mod. Phys. {\bf 71}, 463 (1999)

%---------------------------------------------------------------------

\bibitem{sstoof} Stoof H T C, J. Stat. Phys. {\bf 87}, 1353 (1997)

\bibitem{perez_garcia} Perez-Garcia V M, Michinel H, Cirac J I, Lewenstein M and Zoller P,
Phys. Rev. A {\bf 56}, 1424 (1997)

\bibitem{shi_zheng} Shi H and Zheng W-M, Phys. Rev. A {\bf 55}, 2930 (1997)

\bibitem{ffetter} Fetter A L, J. Low Temp. Phys. {\bf 106}, 643 (1997)

\bibitem{leggett}
Ueda M and Leggett A J, Phys. Rev. Lett. {\bf 80}, 1576 (1998)

\bibitem{parola} Parola A L, Salasnich L and Reatto L, Phys. Rev. A {\bf 57}, R3180 (1998)

\bibitem{kagan}
Kagan Y, Muryshev A E, and Shlyapnikov G V, Phys. Rev. Lett.
{\bf 81}, 933 (1998)

\bibitem{huang}
Eleftheriou E and Huang K, Phys. Rev. A {\bf 61}, 043601
(2000)

\bibitem{adhikaricyl}
Adhikari S K, Phys. Rev. E {\bf 65}, 016703 (2002)

\bibitem{gammal1}
Gammal A, Frederico T and Tomio L, Phys. Rev. A {\bf 64},
055602 (2001); Gammal A, Tomio L and Frederico T, Phys. Rev. A {\bf 66},
043619 (2002)

%\bibitem{fesh} Inouye S, Andrews M R, Stenger J, Miesner H-J, Stamper-Kurn D M and Ketterle W 1998
%Nature {\bf 392} 151 \\
%Roberts J L, Claussen N R, Cornish S L and Wieman C E 2000 Phys. Rev. Lett. {\bf 85} 728 \\
%Robert A et al. 2001 Science {\bf 292} 461

\bibitem{hulet5}
Sackett C A, Stoof H T C, and Hulet R G, Phys. Rev. Lett.
{\bf 80}, 2031 (1998)

\bibitem{hulet6}
Sackett C A, Gerton J M, Welling M, and Hulet R G, Phys.
Rev. Lett. {\bf 82}, 876 (1999)

\bibitem{huletsoliton} Strecker K E, Partridge G B, Truscott A G, and Hulet R G,  Nature 417, 150 (2002);Al Khawaja U,
Stoof H T C, Hulet R G, Strecker K E, and Partridge G B, Phys. Rev. Lett. 89, 200404 (2002)

\bibitem{donley}
Donley E A, Claussen N R, Cornish S L, Roberts J L, Cornell E A, and Wieman C E, Nature {\bf 412}, 295 (2001); Roberts J L,
 Claussen N R, Cornish S L, Donley E A, Cornell E A, and Wieman C E, Phys. Rev. Lett. {\bf 86}, 4211 (2001)

% -----------------------------------------------

\bibitem{duine}
Duine R A and Stoof H T C, Phys. Rev. Lett. {\bf 86}, 2204 (2001)

\bibitem{adhikari2}
Adhikari S K, Phys. Rev. A {\bf 66}, 013611 (2002); S. K. Adhikari, Phys. Rev. A {\bf 66}, 043601 (2002)

\bibitem{santos}
Santos L and Shlyapnikov G V, Phys. Rev. A {\bf 66}, 011602
(2002)

\bibitem{saito}
Saito H and Ueda M, Phys. Rev. A {\bf 65}, 033624 (2002)

\bibitem{savage}
Savage C M, Robins N P and Hope J J, Phys. Rev. A {\bf 67},
014304 (2003)

\bibitem{hollandnjp}
Milstein J N, Menotti C, and Holland M J, New J. Phys. {\bf
5}, 52 (2003)

\bibitem{adhikarijpb}
Adhikari S K, J. Phys. B: At. Mol. Opt. Phys. {\bf 37}, 1185
(2004)

\bibitem{bao}
Bao W, Jaksch D, and Markowich P A, J. Phys. B: At. Mol. Opt.
Phys. {\bf 37}, 329 (2004)

% -----------------------------------------------

\bibitem{mit_dw} Andrews M R, Townsend C G, Miesner H J, Durfee D S, Kurn D M and Ketterle W 1997 Science {\bf 275} 637

\bibitem{magnetic_dw}
Tiecke T G, Kemmann M, Buggle C, Shvarchuck I, von
Klitzing W and Walraven J T M, J. Opt. B:Quantum Semiclass. Opt.
{\bf 5}, S119 (2003)

\bibitem{mit_interferometer_2}
Shin Y, Saba M, Pasquini T A, Ketterle W, Pritchard D E, and
Leanhardt A E, 2004  Phys. Rev. Lett. {\bf 92}, 050405

\bibitem{distillation}
 Shin Y, Saba M, Schirotzek A, Pasquini T A, Leanhardt A E, Pritchard D E, and Ketterle W,
 Phys. Rev. Lett. {\bf 92}, 150401 (2004)

\bibitem{foot}
Thomas N R and Wilson A C, Foot C J, 2002 Phys. Rev. A {\bf 65}, 063406

\bibitem{Two_State_0}
Jack M W, Collett M J and Walls D F, Phys. Rev. A {\bf 54},
R4625 (1996)

\bibitem{Two_State_2} Smerzi A, Fantoni S, Giovanazzi S and Shenoy S R 1997 Phys. Rev. Lett. {\bf 79} 4950

\bibitem{shen} Raghavan S, Smerzi A, Fantoni S, and Shenoy S R 1999 Phys. Rev. A {\bf 59} 620

\bibitem{Two_State_3} Zapata I, Sols F and Leggett A 1998 Phys. Rev. A {\bf 57} R28

\bibitem{Two_State_5} Wu B and Niu Q 2000 Phys. Rev. A {\bf 61} 023402

%\bibitem{leggett2}  Leggett A J 1999, in {\it Proceedings of the 16th
%International Conference on Atomic
%Physics}, Windsor, Ontario, Canada, Aug. 1998, edited by W. E. Baylis and G. F.
%Drake, AIP Conf. Proc. No. 477
%(AIP, Woodbury, New York), pp. 154-169

\bibitem{MF_3} Williams J 2001 Phys. Rev. A {\bf 64} 013610

\bibitem{fant} Giovanazzi S, Smerzi A and Fantoni S 2000 Phys. Rev. Lett. {\bf 84} 4521

\bibitem{sak1} Sakellari E, Leadbeater M, Kylstra N J and Adams C S 2002 Phys. Rev. A {\bf 66} 033612

\bibitem{sak2} Sakellari E, Proukakis N P, Leadbeater M, and Adams C S 2004 New J. Phys. {\bf 6} 94

\bibitem{adhikari}
Adhikari S K, J. Phys. B:At. Mol. Opt. Phys. {\bf 36}, 2943
(2003)

\bibitem{coullet}
Coullet P and Vandenberghe, J. Phys. B:At. Mol. Opt. Phys. {\bf
35}, 1593 (2002)

\bibitem{Loop} Wu B, Diener R B and Niu Q 2002 Phys. Rev. A {\bf 65} 025601 \\
Diakonov D, Jensen L M, Pethick C J and Smith H 2002 Phys. Rev. A {\bf 66} 013604 \\
Mueller E J 2002 Phys. Rev. A {\bf 66} 063603

\bibitem{numrec}
Press W H, Teukolsky S A, Vetterling W T and Flannary B P,
{\it Numerical recipes in FORTRAN : the art of scientific
computing} 2nd Ed. {CUP, Cambridge, 1992}.

\bibitem{number1} Leggett A J and Sols F 1991 Found. Phys. {\bf 21}, 353.

\bibitem{number2} Javaneinen J and Ivanov M Yu 1999 Phys. Rev. A {\bf 60}, 2351.

\bibitem{thermal1} Ruostekoski J and Walls D F 1998 Phys. Rev. A {\bf 58}, R50.

\bibitem{thermal2} Zapata I, Sols F and Leggett A J 1998 Phys. Rev. A {\bf 57}, R28.


\end{thebibliography}
\end{document}